\documentclass[12pt]{article}
\usepackage[utf8]{inputenc}
\usepackage[english]{babel}
\usepackage{adjustbox}
\pdfoutput=1
\usepackage{subfigure}
\usepackage[square,numbers]{natbib}
\usepackage{amsmath}

\title{Emergent Phases in an Extended Bose-Hubbard Model with Three-body On-site Interaction}
\author{
        Henry K. Otobrise \\
        Department of Physics, Lead City University,\\
        Ibadan, Nigeria
            \and
        Kunle Adegoke \\
        Department of Physics and Engineering Physics,\\
        Obafemi Awolowo University,\\
        Ile-Ife, Nigeria
}
\date{\today}
\begin{document}
\maketitle

\begin{abstract}
We study the Extended Bose-Hubbard Model with a three-body onsite interaction. Using an exact diagonalization method and a variational Matrix Product States algorithm, \emph{Alps mps-optim}, we compute and analyse the energy charge gap and finite system correlation functions to obtain the phase boundaries and determine the phases and phase transitions of the model. For \mbox{$\frac{V}{U} = 0.6$} and \mbox{$\frac{W}{U} = 0$}, we observed the emergence of the Haldane insulator (HI) phase at the tip of the \mbox{$\rho = 1$} charge density wave (CDW) lobe and for \mbox{$\frac{V}{U} = 0.6$} and \mbox{$\frac{W}{U} = 1$}, we observed the emergence of the supersolid (SS) phase at the tip of the \mbox{$\rho = 1/2$} charge density wave (CDW) lobe. We also noticed that the effect of the inclusion of both the three-body onsite and nearest neighbour interactions on the phase diagram of the model for filling factor \mbox{$\rho = 2$} is to stabilize the Mott insulator (MI) phase such that the system remains in the MI phase for all values of $\frac{V}{U}$.\\
PACS numbers: \, 03.75.Hh, 03.75.Lm, 3.75.Nt, 67.85.-d, 
\end{abstract}

\section{Introduction}
The Bose Hubbard Model (BHM) includes only a two-body on-site interaction and no inter-site interactions. However, recent experiments with cold atomic gases in optical lattices and Josephson junction arrays (JJAs) have indicated the possibility of on-site interactions beyond two-body \cite{Will10} and long-range inter-site interactions. Recent success with the confinement and manipulation of cold bosonic atoms in optical lattices has led to the possibility of experimental setups where many-body and indeed
long-range interaction terms may appear. One such scenario, proposed by \citet{Buchler07}, considered that polar molecules in
optical lattices are driven by microwave fields and that the system can be described by an effective Hamiltonian where three-body interactions are dominant. It has also been shown that it is possible to control two- and three-body interactions between
molecules in an optical lattice independently.
To study the effects of these interactions, the BHM has to be modified, typically to include interactions such as a three-body on-site interaction and/or a density-density interaction between different sites leading to the appearance of various model variations under the generic name \emph{extended Bose-Hubbard model}.

The earliest of such modified models, which takes into account, additionally, the nearest-neighbour interaction, is
generally termed the \emph{Extended Bose-Hubbard Model} (EBHM) in the literature. Its Hamiltonian is given by
\begin{equation}
\label{ch1:eqnex1} \hat{H}_{EBH} = - t\sum\limits_i
(\hat{a}_i^{\dagger}\hat{a}_{i+1} + \hat{a}_{i+1}^{\dagger}\hat{a}_i) +  \frac{U}{2} \sum_i \hat{n}_i(\hat{n}_i -1) +
V \sum_i \hat{n}_i \hat{n}_{i+1} - \mu \sum_i \hat{n}_i
\end{equation}
or, equivalently
\[
\hat{H}_{EBH} = \hat{H}_{BH} + V \sum_i \hat{n}_i \hat{n}_{i+1}
\]
where $\hat{a}^{\dagger}_i$ and $\hat{a}_i$ are the bosonic creation and
annihilation operators respectively and $ \hat{n}_i = \hat{a}^{\dagger}_i \hat{a}_i$
is the number operator on site $i$, $t$ is the amplitude for the
hopping of bosons between nearest-neighbour pairs of sites $i$, and $i+1$.
$U$ is the on-site two-body interaction amplitude, $V$ is the nearest-neighbour interaction parameter while $\mu$ is the chemical potential.
$\hat{H}_{BH}$ is the Hamiltonian of the Bose-Hubbard model (i.e. when $V = 0$).

Although the essential qualitative features of the BHM phase diagram, such as the existence of Mott insulator (MI) lobes at integer filling surrounded by Superfluid (SF) phase were first worked out by \citet{Fisher89} using a mean-field approach, the study of the BHM's SF-MI transition continues to attract much attention, with a large body of work, using various methods such as the exact diagonalization (ED), DMRG, perturbation methods and QMC, etc, attempting to enhance the quantitative understanding of its structure \citep{Fisher89,Elstner99,Kuhner98,Kuhner00,Zakrzewski08,Ejima11}.

The phase diagram of the standard EBHM shows, in addition to the conventional phases observed in the BHM, the existence of novel quantum phases such as charge density wave (CDW) \citep{Kuhner98}, checkerboard solids, supersolid (SS) phases \citep{Batrouni13}, exotic Haldane insulators \citep{Dalla06,Rossini12}, pair-correlated superfluid phases and more. Although the EBHM has been extensively researched for over two decades now, its phase diagram is yet to be fully determined over all the physically relevant parameters of its Hamiltonian. For instance, it had been generally conceded that the SS phase cannot exist in a one-dimensional BHM (except in the case of doping above half filling), however, the publication by \citet{Batrouni13}, predicts a SS phase for the one-dimensional EBHM at filling factor of $\rho=2$. Hence, the need to further study the impact of long-range interaction on the phase diagram of the BHM.

Also, an extended variation of the BHM that includes a three-body on-site interaction has, for sometime now, been the focus of active research \citep{Chen08,Sowinski12a,Sowinski14,Varma14}.

Interestingly, several studies have shown that the presence of long-range interactions noticeably enriches the phase diagram of the BHM with the emergence and stabilization of certain exotic quantum phases such as the bosonic Haldane insulator (HI) phase \citep{Dalla06,Berg08,Iskin11}.
\section{The Model Hamiltonian}
To further understand the impact of long-range interactions on the phase diagram of the BHM, we study in this work a one-dimensional model variation of the BHM that includes a three-body on-site and nearest neighbour interactions with a tight-binding Hamiltonian given by
\begin{align}
\label{ch1:eqnm1}
\nonumber
\hat{H}_{MH} & = - t\sum\limits_i
(\hat{a}_i^{\dagger}\hat{a}_{i+1} + \hat{a}_{i+1}^{\dagger}\hat{a}_i) +  \frac{U}{2} \sum_i \hat{n}_i(\hat{n}_i -1) +
\frac{W}{6} \sum_i \hat{n}_i(\hat{n}_i -1)(\hat{n}_i -2) \\
& \quad + V \sum_i \hat{n}_i \hat{n}_{i+1} - \mu \sum_i \hat{n}_i
\end{align}
or, equivalently
\begin{equation}
\hat{H}_{MH} = \hat{H}_{BH} + \frac{W}{6} \sum_i \hat{n}_i (\hat{n}_i
-1)(\hat{n}_i -2)+ V \sum_i \hat{n}_i \hat{n}_{i+1}
\end{equation}
where $W$ is the three-body interaction amplitude, $V$ is the
nearest-neighbour repulsion amplitude and $\mu$ is the chemical potential.

\section{Previous Works}\label{previous work}
 
Recent successes with the confinement and
manipulation of cold bosonic atoms in optical lattices has led to
the possibility of experimental setups where many-body and indeed
long-range interaction terms may appear. One such scenario was
proposed by \cite{Buchler07} that polar molecules in
optical lattices are driven by microwave fields and that the system can be described by an effective Hamiltonian where
three-body interactions play the dominant role. It has also been shown that it
is possible to control two- and three-body interactions between
molecules in an optical lattice independently.
These long range interactions can be viewed, at least qualitatively, as perturbations
to the standard BHM so that when the perturbation vanishes, we
recover the phase diagram of the BHM. Specifically then, we look at how the addition of long range interactions modify
the phase diagram of the standard BHM.
We review below the literature of two model variations of the BHM.
\subsection{Bose-Hubbard Model with Three-Body On-site Interaction}
\label{mbh.4} 
The first variation, in contrast to the
standard BHM where only two-body on-site interaction is
present, consists additionally of a three-body on-site (i.e. local )
interaction term. \newline
The Hamiltonian, $\hat{H}_{BH3b}$ of this model is given as,
\begin{equation}
\label{ch2:eqn53}
\hat{H}_{BH3b} = -t \sum\limits_{i} \left(  \hat{a}^{\dagger}_i \hat{a}_{i+1} + \hat{a}_{i+1}^{\dagger}\hat{a}_i
\right) +\frac{U}{2} \sum \limits_{i}\hat{n}_i(\hat{n}_i-1)+\frac{W}{6} \sum \limits_{i}\hat{n}_i(\hat{n}_i-1)(\hat{n}_i-2) - \mu\sum \limits_{i}\hat{n}_i
\end{equation} 
where $W>0$ is the three-body repulsive interaction amplitude. Or,
\[\hat{H}_{BH3b} = \hat{H}_{BH}+\hat{H}_{3bINT} \, , \]
where $\hat{H}_{3bINT}= -\frac{W}{6} \sum \limits_i
\hat{n}_i( \hat{n}_i-1)(\hat{n}_i-2)$ is the three-body interaction term.

This model has been studied by various researchers using ED
\citep{Sowinski12}, perturbative mean-field theory and the Gutzwiller
variational \emph{ansatz} \citep{Sowinski14}, DMRG \citep{Sowinski14,Valencia11} and the strong-coupling perturbation expansion technique \cite{Varma14}.
The general results from these studies show that like the BHM, the phase diagram of this model exhibits the MI lobes for integer filling surrounded by an SF phase. However, while the first MI lobe is apparently unaffected, there is a non-trivial effect of the three-body
interaction on the Mott lobes corresponding to filling factors higher than one. The primary effect of increasing $W$, compared with
the standard BHM is the increase of MI lobe widths and increased stability of the MI lobes with respect to hopping strength.
 The non-trivial effect of the three-body Interaction, $W$, on the Second MI Lobe a typical phase diagram of the model variation with $W/U = 1$ shows that the BKT point for this lobe for $W/U = 0$ is at $t/U = 0.3$ while the BKT point for $W/U = 1$ is located at $t/U = 0.38$. \citep{Sowinski12,Sowinski14,Sowinski14b,Varma14}.
 
\subsection{The One-Dimensional Extended Bose-Hubbard Model}
\label{mbh.3} 
Recent progress in cooling and condensing Chromium atoms which, unlike alkali atoms,
have a large dipole moment which leads to long-range dipolar
interactions suggests that the extension to the BHM to include a nearest-neighbour interaction is both
experimentally and theoretically relevant \citep{Iskin11}.
We now review the the physics and phase diagram of the BHM with additional nearest-neighbour interaction between
bosons --- the \emph{Extended Bose-Hubbard model} (EBHM).
The Hamiltonian of the one-dimensional EBHM \citep{Kuhner00} is given by
\begin{equation}
\label{bh.ebh1} \hat{H}_{EBH} = - t\sum\limits_i
(\hat{a}_i^{\dagger}\hat{a}_{i+1} + \hat{a}_{i+1}^{\dagger}\hat{a}_i) + \frac{U}{2} \sum_i
{\hat{n}_i}(\hat{n}_i -1) + V \sum \limits_{i}\hat{n}_i
\hat{n}_{i+1}-\mu\sum \limits_{i }\hat{n}_i ,
\end{equation}
or
\begin{equation}
\label{bh.ebh2} \hat{H}_{EBH} = \hat{H}_{BH} + V \sum
\limits_{i}\hat{n}_i \hat{n}_{i+1},
\end{equation}
where $ V $ is the nearest-neighbour repulsive interaction parameter (i.e. $ V >0 $).
For the case $ V \,= \, 0 $, Equation~\eqref{bh.ebh2} reduces to that of the standard BHM.
\subsubsection{Phase Diagram of the One-dimensional Extended Bose-Hubbard Model}
The presence of long range interactions significantly enriches the
phase diagram of the standard BHM. It is by now well
established that the presence of nearest-neighbour repulsive
interaction leads to the emergence of several new phases in addition
to the SF and MI phases present in the
BHM. Several studies have been conducted to fully determine and
characterize the phase diagram of this extended model in
one-dimension \citep{Kuhner98,Kuhner00,Dalla06,Berg08,Batrouni13}. We
now review the results of some of these studies.

In two early works on this model \citep{Kuhner98,Kuhner00}, using the
DMRG method on systems of up to 512 lattice sites and with filling
factor ranging from $1/2$ to $3/2$, it was found that in the presence
of nearest-neighbour repulsive interaction a new insulator phase
emerges at half-integer densities. This phase, the charge density
wave (CDW) phase, is characterized by periodic density fluctuation
around the average filling. The CDW has a wavelength of two sites
and like the MI phase it has an excitation gap and is
incompressible.

\begin{figure}[!hbp]
\begin{center}
\fbox{
\includegraphics[scale = 0.6]{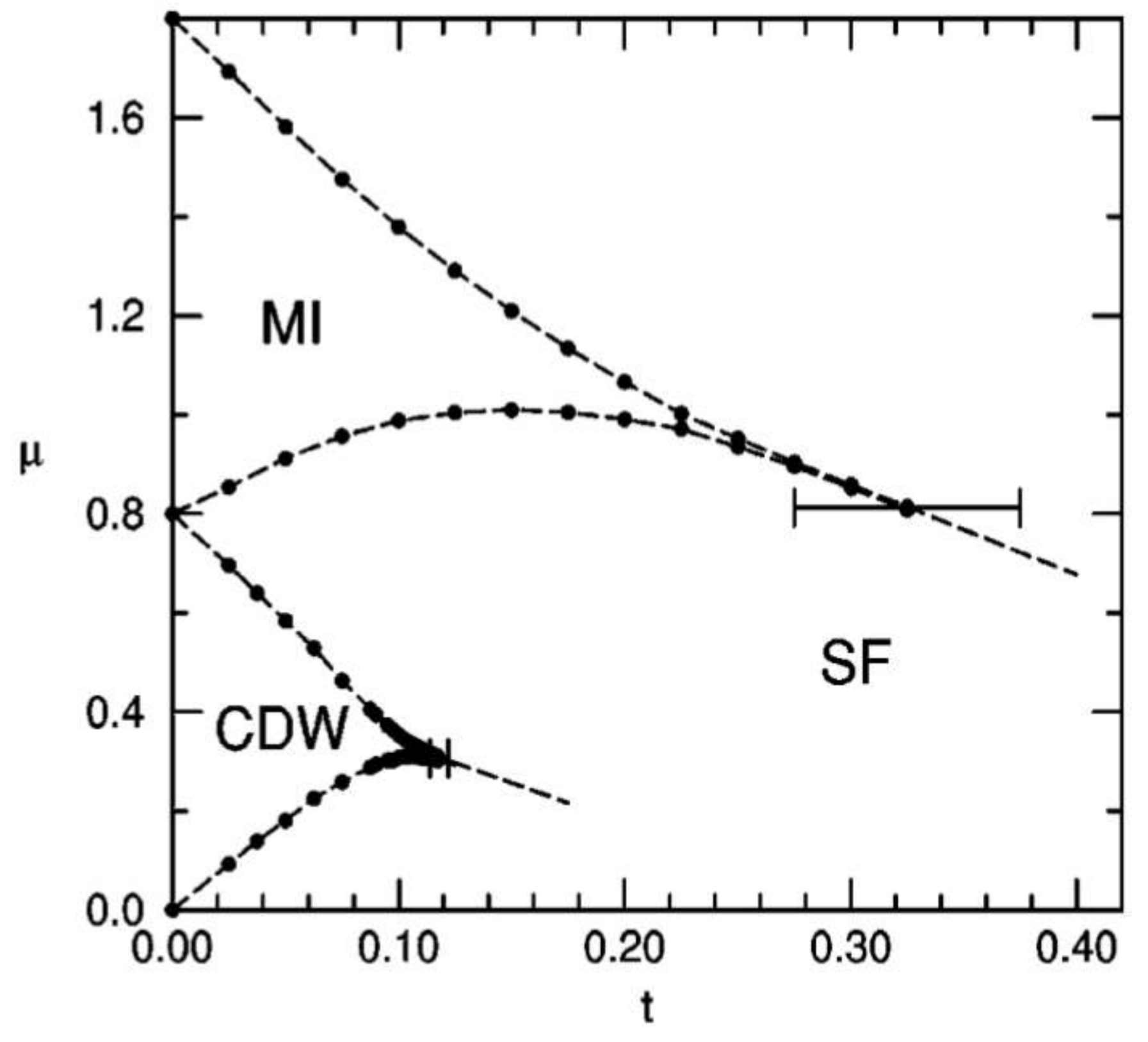}}%
\caption{\label{ch2:kuhner1} Phase Diagram of the One-dimensional
EBHM \mbox{(U = 1,V = 0.4)}. The Phases are: MI with Filling One,
CDW with Filling One-half and SF Phases. Circles Represent DMRG Results. Adapted from
\citet{Kuhner98}.}
\end{center}
\end{figure}

The CDW phase was found to have similar shape as the MI lobes in the
standard BHM with BKT transition at the tip of the lobe.
The CDW lobe like the MI lobe also displays the re-entrance phase
transition from SF to the insulating phases. They found the
BKT transitions for this model at \mbox{$t_c^{MI} \, = \, 0.404\pm0.002$}
for the MI phase and at \mbox{$t_c^{CDW} \, = \, 0.125\pm0.003$} for
the CDW phase \citep{Kuhner00}. Figure~\ref{ch2:kuhner1} shows the
ground state phase diagram of the one-dimensional EBHM
in the ($ \mu $, $t$)-plane as obtained by \cite{Kuhner00}
(the value of $U$ is set to 1).
Outside the CDW phase, no other new phase, including the supersolid phase (SS), was reported in these works.

\clearpage

In a major work, \citet{Rossini12} also studied the EBHM using the DMRG algorithm with open boundary conditions. By performing an up-to-date finite-size scaling with systems of sizes up to $L=400$ sites they too, in agreement with the previous observations above, found the conventional SF and MI phases as well as the CDW. 
However, they found an additional new insulating phase which displays a long-range string order. This is the same Haldane insulator originally predicted by \citet{Dalla06}. The Haldane insulator (HI) phase is a topological quantum phase: it has no local order parameter and hence is not describable within the phenomenological Landau paradigm of condensed matter \citep{Dalla06}.

In another work, \citet{Batrouni13} used the stochastic Green function quantum Monte Carlo (QMC) method and the DMRG algorithm to study this model for filling factors of $\rho \leq 3$ and various ratios of the nearest neighbour interaction $V$ to the on-site interaction $U$ but reported for the special case of $V/U \,=\, 3/4$ which they found favours CDW phase over MI phases at commensurate fillings when $U$ is large.
Their results showed, for the parameter regime considered, the presence of the conventional SF and the emergence of three new phases --- the CDW phase, the HI phase, and surprisingly a SS phase.

\section{Methods}\label{methods}
To obtain the quantum phase diagram of our model Hamiltonian, we studied the Hamiltonian in detail over its physically relevant parameters using an exact diagonalization scheme and a variational (DMRG-like) matrix product states (MPS) algorithm. The phase boundaries were obtained using an exact diagonalization scheme and finite scaling method with a maximum system size of 12 sites
with no limitation placed on site occupancy. On the other hand, the ground state properties that were used to characterise the phases and phase transitions of the model were obtained using the \emph{mps-optim} code which is the Alps implementation of the variational MPS algorithm \cite{Dolfi14}. Here system size ranged from 32 to 100 sites, while site occupancy was limited to a maximum of five particles per site. Where the ground state properties in the thermodynamic limit are not easily deducible directly from the results of the finite system, finite-size scaling methods were then employed to obtain the relevant ground state properties in the thermodynamic limit.
With the order parameters calculated, the phases and phase transitions were then determined using Table 1.
   
\subsection{Properties of the Ground State of the Model Hamiltonian}
\label{ch3:ED-prop} To characterize the phase diagram of our model
Hamiltonian and its variations, we define several quantities that we use for that purpose in this study:
The charge gap which is defined by
\begin{equation}
\label{ch3:cgp}
\Delta_c \, = \, \mu_L (N)- \mu_L (N-1) ,
\end{equation}
where $ \mu_L(N) \, = \, E_{grd}(L,N+1) \, - \, E_{grd}(L,N)$
defines the the chemical potential, and $E_{grd}(L,N)$ the ground
state energy for a system of $N$ bosons on $L$ sites.

The neutral gap which is defined by
\begin{equation}
\label{ch3:ngp}
\Delta_n \, = \, E_0 (L,N)- E_1 (L,N) ,
\end{equation}
where $E_0 (L,N)$ is the ground state energy and $E_1 (L,N)$ the first excited state energy for a system of $N$ bosons on $L$ sites.

The superfluid order correlation function $C_{SF}$ is defined by
\begin{equation}
\label{ch3:sf} C_{SF} (r) \, = \, \langle \hat{a}_j^{\dagger}
\hat{a}_{j+r} \rangle \ ;
\end{equation}

The charge density wave correlation function $C_{CDW}$ is defined by
\begin{equation}
\label{ch3:cdw1} C_{CDW} (r) \, = \, (-1)^r \langle \delta \hat{n}_j
\delta \hat{n}_r \rangle,
\end{equation}

The charge density wave structure factor $S_{k}$ \citep{Batrouni13} is defined by the Fourier transform (F.T) of the density-density correlation function
\begin{equation}
\label{ch3:cdw2} S_{k} \, = \, \frac{1}{L} \sum \limits_{r=0}^{L-1}
e^{ikr} \langle \hat{n}_0 \hat{n}_r \rangle,
\end{equation}

The momentum distribution $n_k$ is defined by
\begin{equation}
\label{ch3:nk} n_{k} \, = \, \sum \limits_{r=0}^{L-1} e^{ikr}
\langle \hat{a}^{\dagger}_0 \hat{a}_r \rangle,
\end{equation}

The expectation value of string operator is defined by
\begin{equation}
\label{ch3.sops} {\mathcal O}_s (|i-j|) \, = \, \left \langle \delta
\hat{n}_i e^{i\theta \sum_{k=i}^j \delta \hat{n}_k} \delta \hat{n}_j
\right \rangle
\end{equation}

The expectation value of parity operator is defined by
\begin{equation}
\label{ch3.pops} {\mathcal O}_p (|i-j|) \, = \, \left \langle
e^{i\theta \sum_{k=i}^j \delta \hat{n}_k} \right \rangle
\end{equation}
where $\hat{a}_j^{\dagger} \, (\hat{a}_j)$ creates (annihilates) a
boson on site $j$, $\delta \hat{n_i} \, = \, \hat{n_i} - \rho I$, $k \, = \, \pi$, $\theta \, = \, \pi$ 
for the cases $\rho$ = 1, 1/2  and $\theta \, = \, \pi /2$ for the cases $\rho$ = 3/2, 2.

\begin{table}[!htp]
\centering
\vspace{2.0in}
\caption{\label{ch3:table1} Classification of Phases Based on Order Parameters.}
\begin{tabular}{|l|l|l|l|l|l}
\cline{1-5}
Phase & $C_{SF}(\infty)$ & $C_{CD}(\infty)$ & ${\mathcal{O}}_s(\infty)$ & ${\mathcal{O}}_p(\infty)$ & \\
\cline{1-5}
SF  & $ \ne \, 0$ & 0 & 0 & 0 & \\
\cline{1-5}
MI & 0 & 0 & 0  & $\ne \, 0$ & \\
\cline{1-5}
HI & 0 & 0 & $\ne \, 0$ &  0 & \\
\cline{1-5}
CDW & 0 & $\ne \, 0$ & $\ne \, 0$ & $\ne \, 0$ &  \\
\cline{1-5}
SS &$ \ne \, 0 $& $\ne \, 0$ & $\ne \, 0$ & $\ne \, 0$ & \\
\cline{1-5}
\end{tabular}
\end{table}

\section{Results}\label{results}
We investigated the Model Hamiltonian for parameter combinations over the following ranges: $ W/U = [0, 0.5, 1.0]$ and $ V/U = [0.2, 0.4, 0.5, 0.6]$. We now present the results of our investigations.

The system was found to be unstable for values of $V/U > 0.6$ hence, results for these are not reported. Thus, the largest value of the nearest-neighbour interaction reported in this work is $V/U$ = 0.6. 
For $0 \, < \, V/U \, \leq \, 0.4$ and for all values of $W/U$, remains essentially similar to that of the EBHM with the phase diagrams showing no new phases outside the MI lobes for commensurate filling factors, CDW lobes for half-integer filling factors surrounded by an SF phase. Although the inclusion of finite $W$ in the EBHM does not affect the phases present but as in the BHM, the second MI lobe is significantly enlarged and the BKT point extended further into the surrounding SF phase. The BKT point for the second MI lobe for the system with $W$ = 0 and $V$ = 0.4 was found at \mbox{$t_c$ = 0.29$\pm$0.01} while that for the system with $W$ = 1.0 and $V$ = 0.4 was found at \mbox{$t_c$ = 0.45$\pm$0.01.}
However, for $V/U \, > \, 0.5$ and $W/U = 0$, we observed a significant change to the phase diagram: all the lobes, for integer filling as well as half-integer filling, are in the CDW phase surrounded by an SF phase. The previously MI lobes for integer filling are now in the CDW phase. With these findings,our detailed report, is therefore, limited to two model variations as follows:
\subsection{Case $W \, = \, 0$; $V \, = \,0.6$}
We investigate this model further to have a better understanding of its phases and phase transitions. We consider the model for system sizes L = 32, 64 and 100 at filling factors $\rho$ = 1/2, 1, 3/2, and 2.
 
We start with the system at unit filling factor $\rho$ = 1. Using a finite size scaling analysis, we have that $C_{SF}(\infty)$ = 0 for values of \mbox{$t \, \leq \, 0.22$, $C_{CDW}(\infty)$} vanishes for $t \, > \, 0.16$, ${\mathcal{O}}_p(\infty)$ also vanishes for $t \, > \, 0.16$, while ${\mathcal{O}}_s(\infty)$ vanishes for $t \, > \, 0.28$. Using the criteria defined in Table~\ref{ch3:table1}, we conclude that the lobe in the model's phase diagram is in a CDW phase for $t \, < \, 0.16$ and in the HI phase for \mbox{$0.16 \, < \, t \, \leq \, 0.22$}. The surrounding phase is SF.

We next consider the system for filling factor $\rho$ = 2. From finite size scaling analysis, we have $C_{SF}(\infty) \, = \, 0$ for values of $t \, \leq \, 0.21$, we estimate that $C_{CDW}(\infty)$ = 0 for $t \, \geq \, 0.21$, ${\mathcal{O}}_p(\infty)$ = 0, for values of $t \, \geq \, 0.22$ and ${\mathcal{O}}_s(\infty)$ = 0 for values of $t \, \geq \, 0.21$. From Table~\ref{ch3:table1}, we conclude that the $\rho$ = 2 lobe in the model's phase diagram is a CDW lobe while the surrounding phase is the SF phase.

For half filling factor (i.e. $\rho$ = 1/2): Finite size scaling yields $C_{SF}(\infty) \, = \, 0$ for values of $t \, \leq \, 0.19$, $C_{CDW}(\infty)$ vanishes for $t \, > 0.30$. For ${\mathcal{O}}_p$ and ${\mathcal{O}}_s$: both parameters are finite over the range of $t$ considered in the study. From Table~\ref{ch3:table1}, it is clear that the $\rho$ = 1/2 lobe in the model's phase diagram is a CDW phase while the surrounding phase is the SF phase.

For filling factor $\rho$ = 3/2: Finite size scaling analysis yields $C_{SF}(\infty) \, = \, 0$ for values of \mbox{$t \, \leq \, 0.18$}, $C_{CDW}(\infty)$ vanishes for \mbox{$t \, > 0.18$}, ${\mathcal{O}}_p(\infty)$ vanishes for values of \mbox{$t \, \geq \, 0.19$} and \mbox{${\mathcal{O}}_s(\infty) \, > \, 0.0$} for all $t$. Clearly, from Table~\ref{ch3:table1}, the $\rho$ = 3/2 lobe in the model's phase diagram is a CDW phase while the surrounding phase is the SF phase.

In addition to the phases, we also estimated the BKT points for the EBHM with $V$ = 0.6: for the insulator lobe at half filling, the phase is CDW, with the BKT point at \mbox{$t_c = 0.19\pm0.01$}. For the lobe at unit filling, the phase is CDW/HI with the BKT point at \mbox{$t_c$ = 0.22$\pm$0.01}; for the lobe at filling \mbox{$\rho = 3/2$}, the phase is CDW with the BKT point at $t_c = 0.18 \pm0.01$ and for the lobe at filling $\rho = 2$, the phase is also CDW with  the BKT point at \mbox{$t_c = 0.21 \pm0.01$}. 

\clearpage
\begin{figure}[!htp]
\begin{center}
\fbox{\includegraphics[scale = 0.25]{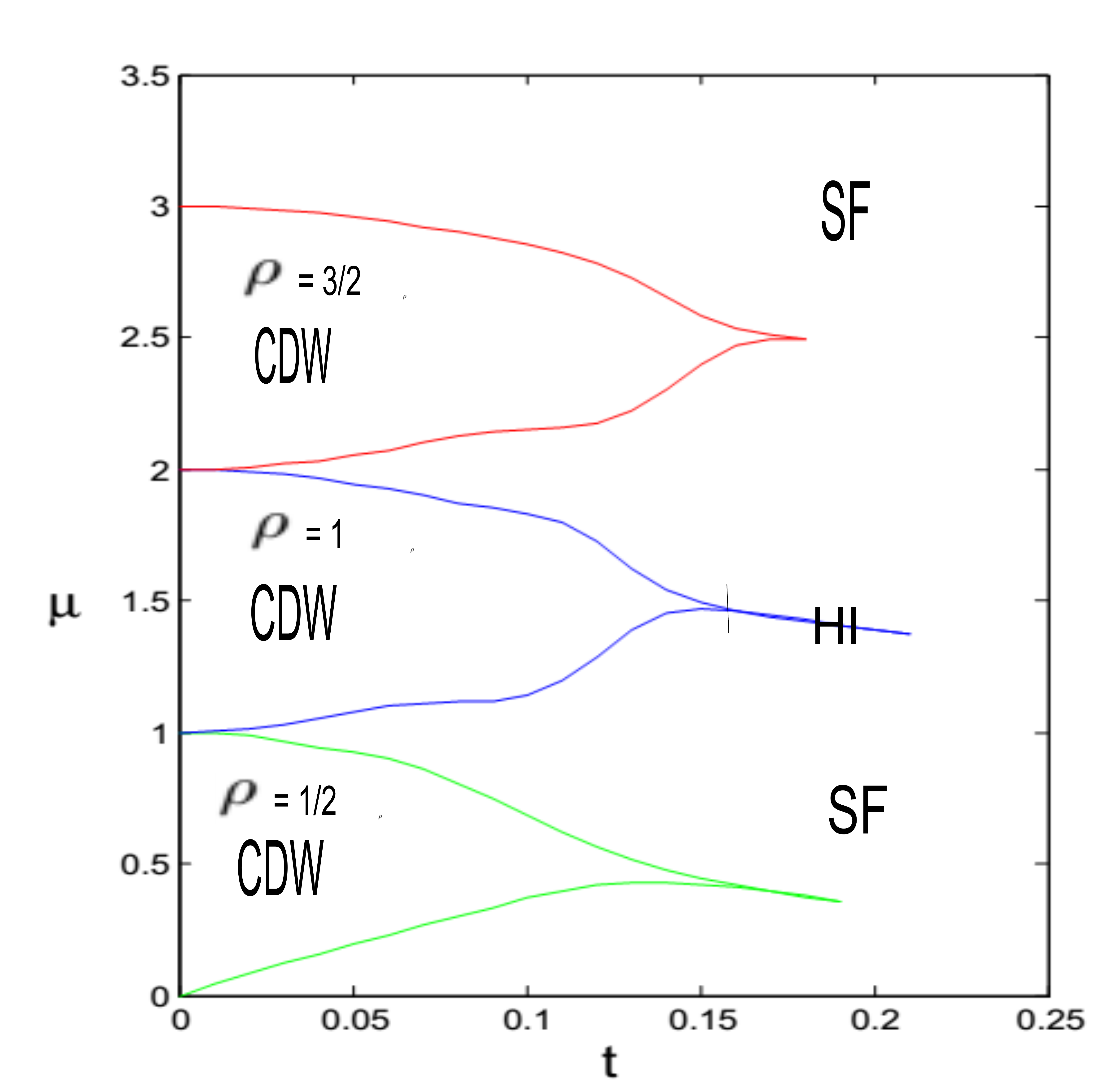}}%
\caption{\label{Ch4:pd060} Phase Diagram of the One-dimensional EBHM \mbox{($U = 1.0$, $W\, = \, 0$, $V = 0.6$)}. Note the HI Phase at the Tip of the \mbox{$\rho = 1$} CDW Lobe with the Vertical Line Marking the CDW-HI Phase Boundary. The Phase Boundary of the \mbox{$\rho = 2$} CDW Phase from the ED Study is not Stable and hence not Shown in the Figure.}
\end{center}
\end{figure} 
The phase diagram, in the \mbox{($\mu$, $t$)}-plane, for this EBHM is shown in Figure~\ref{Ch4:pd060}.
 
\clearpage
\begin{figure}[!htp]
\label{Ch4:ebhwv622}
\centering
\adjustbox{stack,fbox}{
\subfigure[]{\label{Ch4:ebhwv62b}
\includegraphics[height=3.0in,width=5.0in,clip=true]{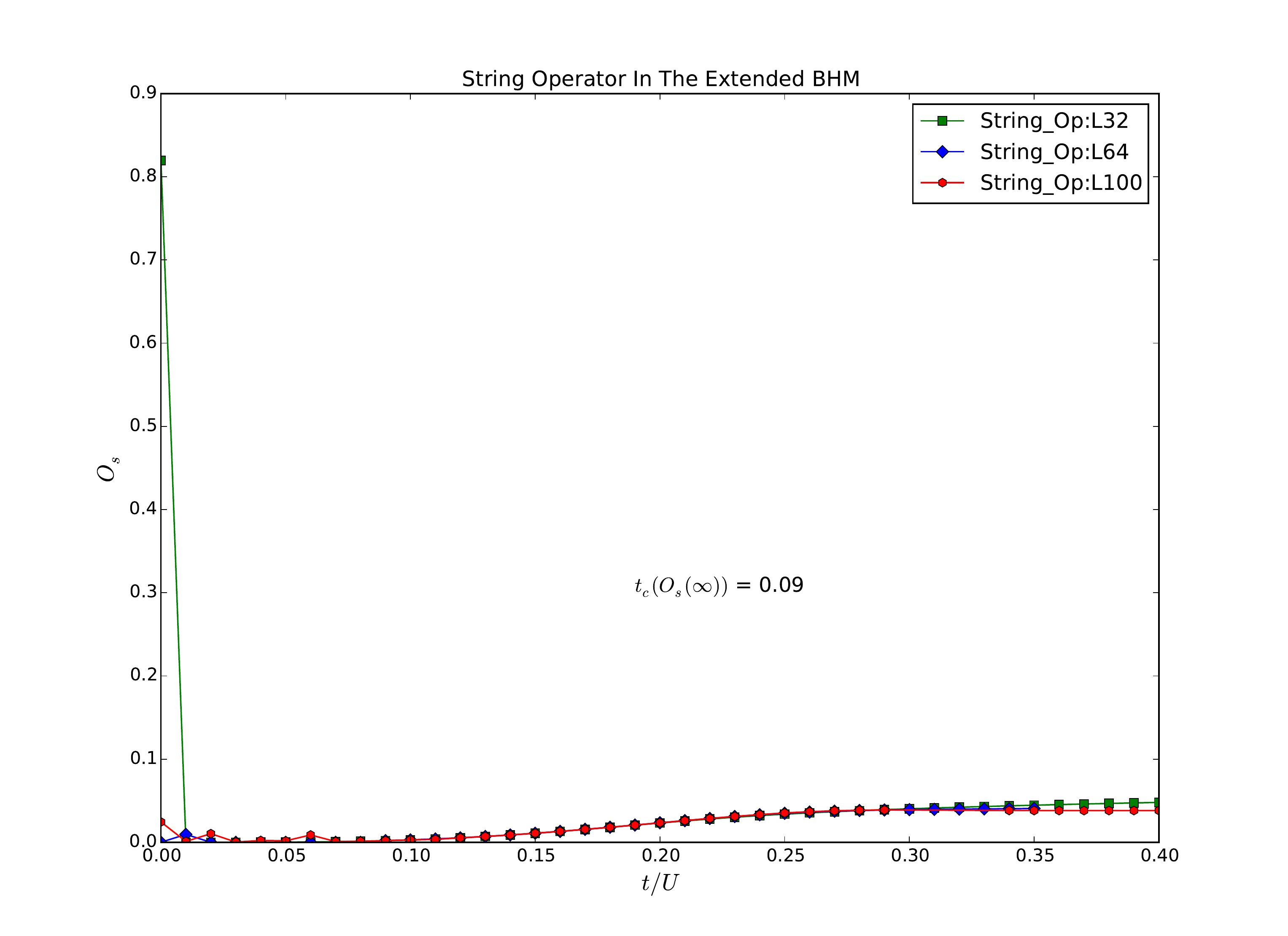}}\\
\subfigure[]{\label{Ch4:ebhwv62d}
\includegraphics[height=3.0in,width=5.0in,clip=true]{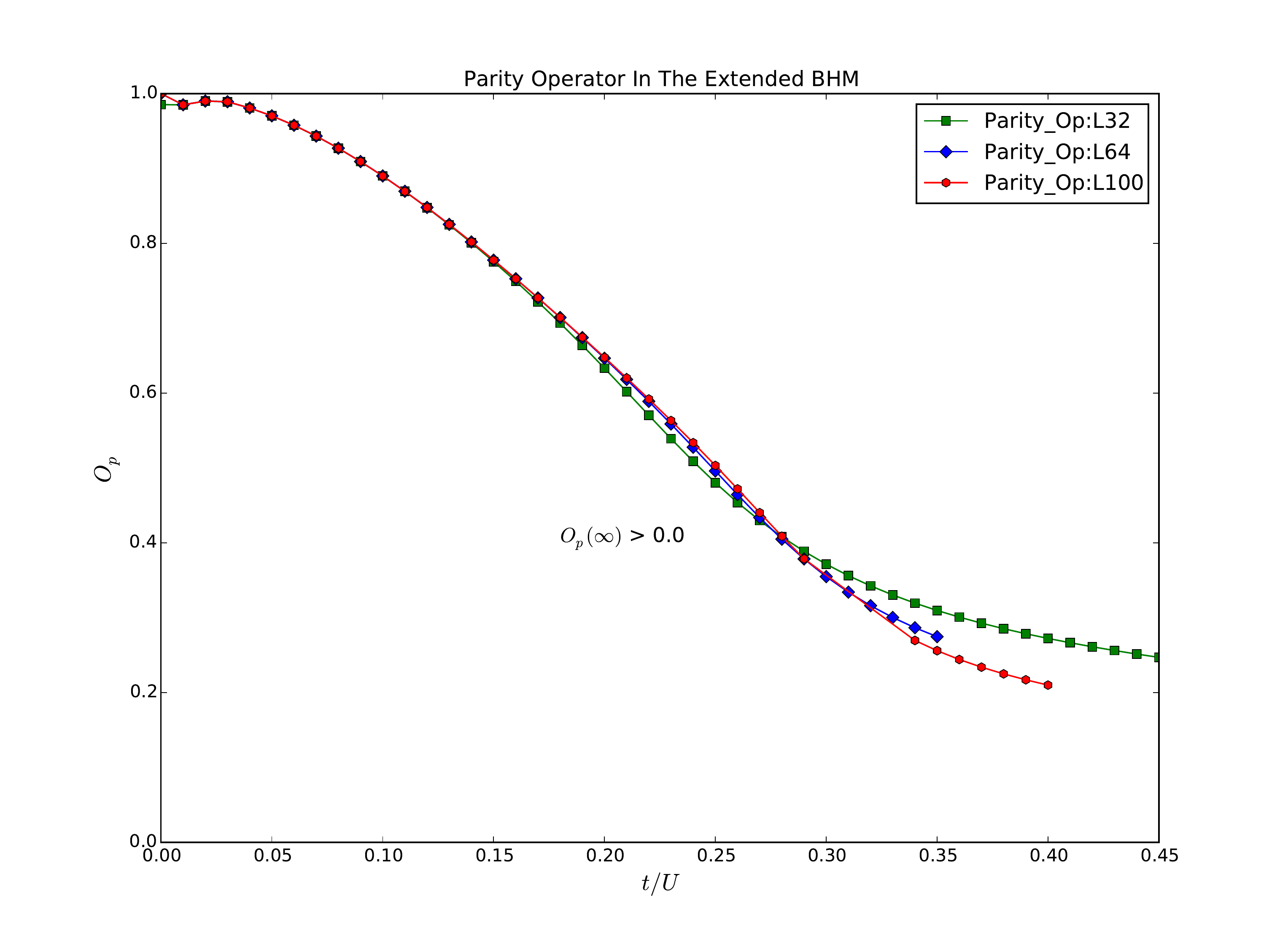}}
}%
\caption{\label{Ch4:ebhwv62}: Plots of (a) \mbox{${\mathcal{O}}_s$}, and (b) \mbox{${\mathcal{O}}_p \,  vs. \, t/U$} for the One-dimensional EBHM \mbox{($W = 1.0$, $V = 0.6$)} for Filling Factor $\rho = 2.$}
\end{figure}

\subsection{Case $W \, = \,1.0$; $V \, = \,0.6$}
\label{Ch4:vmps_8}
An interesting feature of the last model studied, where $V$ = 0.6, is the emergence of the HI phase at unit filling.
\newline From previous studies, the three-body on-site interaction is known to stabilize the second MI lobe in the phase diagram of the BHM. We now investigate the effect of this interaction in a model that originally exhibits the CDW, HI and SF phases.   

We start with the system at unit filling factor $\rho$ = 1. Finite size scaling analysis yields $C_{SF}(\infty)$
= 0 for values of \mbox{$t \, \leq \, 0.18$}, $C_{CDW}(\infty)$ vanishes for $t \, > \, 0.16$,
${\mathcal{O}}_p(\infty)$ vanishes for \mbox{$t \, \geq \, 0.30$} while \mbox{${\mathcal{O}}_s(\infty)$ = 0} for values of \mbox{$t \, \geq \, 0.17$}. Using the criteria defined in Table~\ref{ch3:table1}, we conclude that the lobe in the model's phase diagram is in CDW phase for \mbox{$t \, \leq \, 0.16$} and in the HI phase for \mbox{$0.16 \, < t \, \leq \, 0.18$} while the surrounding phase is the SF. We note that the impact of the three-body on-site repulsive interaction in this model is to stabilize the CDW phase in preference to the emerging HI phase. 

We now consider the system for filling factor $\rho$ = 2. From finite size scaling analysis, we have that $C_{SF}(\infty) \, = \, 0$ for values of $t \, \leq \, 0.38$. We note that $C_{CDW}(\infty)$ = 0 and \mbox{${\mathcal{O}}_p(\infty) \, > \, 0.0$} for all $t$, respectively. While ${\mathcal{O}}_s(\infty)$ = 0 for values of $t \, \leq \, 0.09$. From Table~\ref{ch3:table1} and given that the $C_{CDW}(\infty)$ = 0 for all $t$, we conclude that the $\rho$ = 2 lobe in the model's phase diagram is MI phase while the surrounding phase is the SF phase. Figure~\ref{Ch4:ebhwv622} shows the plots for ${\mathcal{O}}_p(\infty)$ and ${\mathcal{O}}_s(\infty)$ against $t/U$.
An interesting result: the introduction of the three-body on-site interaction in this model actually favours the existence of the MI phase over the CDW phase. 

For filling factor $\rho$ = 1/2: Using a finite size analysis, we have that \mbox{$C_{SF}(\infty) \, = \, 0$} for values of $t \, \leq \, 0.19$, $C_{CDW}(\infty)$ = 0.0 for $t \, > \, 0.25$ and \mbox{${\mathcal{O}}_s(\infty) \, > \, 0.0$} for all $t$. Similarly, \mbox{${\mathcal{O}}_p(\infty) \, > \, 0.0$} for all $t$. From Table~\ref{ch3:table1}, we conclude that the $\rho$ = 1/2 lobe in the model's phase diagram is in the CDW phase. Interestingly, we note the emergence of a new phase --- the SS phase --- between the CDW lobe and surrounding SF phase. The SS phase is found between $0.19 \, < t \, \leq \, 0.25$.

\begin{figure}[!htp]
\begin{center}
\fbox{\includegraphics[scale=0.3]{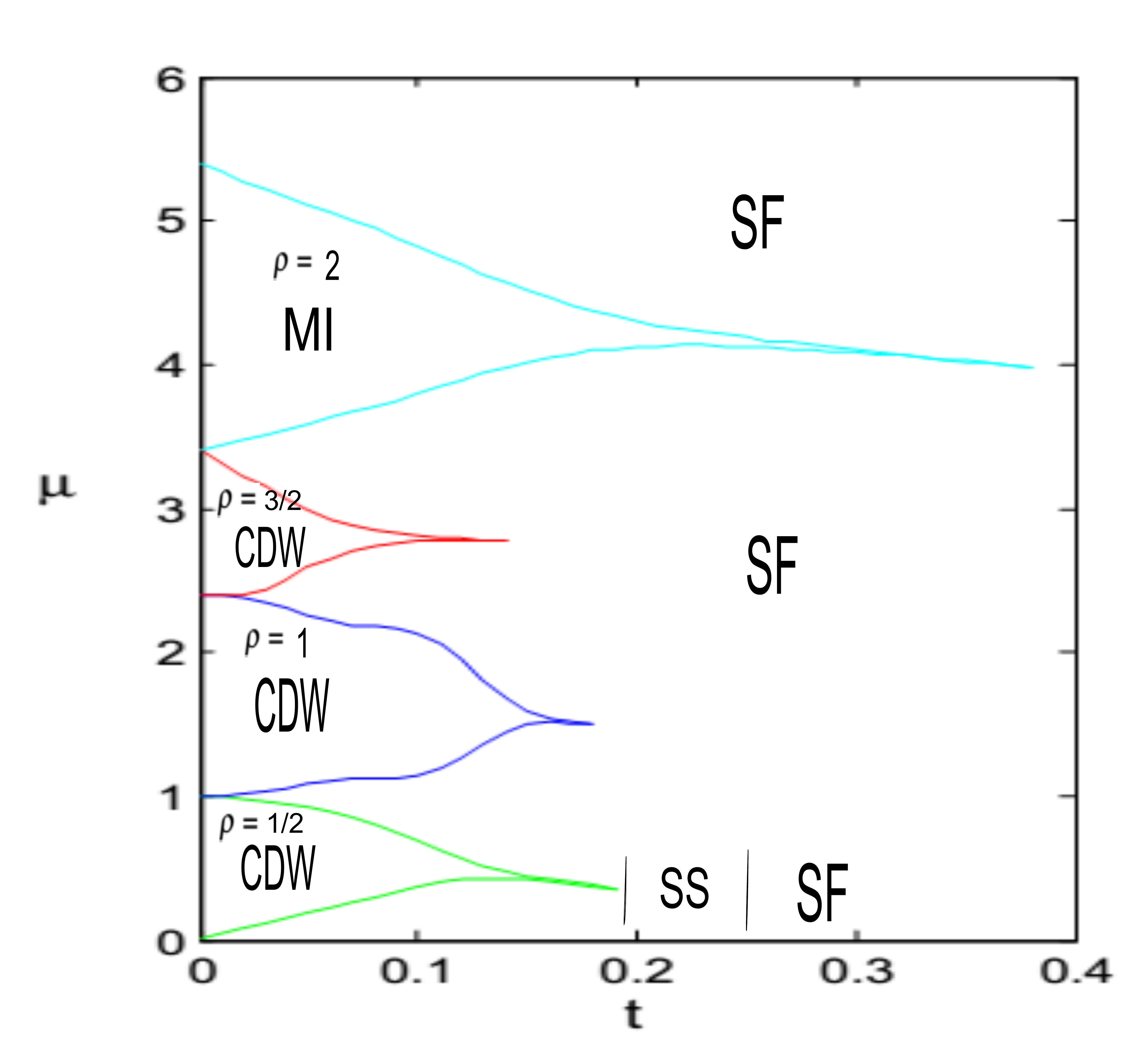}}%
\caption{\label{Ch4:pd160} Phase Diagram of the One-dimensional
EBHM \mbox{($U = 1.0$, $W\, = \, 1.0$, $V = 0.6$)}. The SS Phase is Shown in Between the Vertical Lines at \mbox{$t = 0.19$ and $0.25.$}}
\end{center}
\end{figure}

For filling factor $\rho$ = 3/2: Using a finite size scaling \emph{ansatz}, we note that $C_{SF}(\infty) \, = \, 0$ for values of $t \, \leq \, 0.15$ and estimate that $C_{CDW}(\infty)$ = 0.0 for $t \, > \, 0.15$. We also estimate that ${\mathcal{O}}_p(\infty) \, > \, 0.0$ for all $t$ and ${\mathcal{O}}_s(\infty)$ is likewise finite for all $t$. From Table~\ref{ch3:table1}, we conclude that the $\rho$ = 3/2 lobe in the model's phase diagram is in a CDW phase while the surrounding phase is the SF phase.
Figure~\ref{Ch4:pd160} shows the system's phase diagram, in the ($t$, $\mu$)-plane.

\clearpage

\section{Conclusions}\label{conclusions}

We obtained the phase diagrams of a one-dimensional BHM extended to include a three-body on-site interaction and a nearest neighbour interaction using two numerical methods \--- the exact diagonalisation method implemented in Matlab by the authors and the \mbox{Alps\_mps\_optim} --- from the Alps project implementation of the variational matrix product state algorithm.

We studied system sizes of up to 100 sites for parameter combinations of (a) $W$ = [0, 1.0], $V$ = 0.4 and (b) $W$ = [0, 1.0], $V$ = 0.6.

We observed the emergence of the Haldane insulator (HI) phase at the tip of the mainly CDW lobe at unit filling for the system with $W$ = 1.0 and $V$ = 0.6.
However, the presence of the three-body on-site interaction($W = 1.0$) was found to stabilize the MI phase for the EBHM at filling factor $\rho$ = 2 where we found the $\rho$ = 2 lobe for the system with $W$ = 1.0 and $V$ = 0.6 to be fully Mott insulating as against the situation where it was fully in the CDW phase for $W$ = 0 and $V$ = 0.6. It also favours the emergence of the much elusive SS phase at half filling, while destroying the emerging HI phase at unit filling.

\bibliographystyle{apalike}
\bibliography{BoseHubbardModel}
\end{document}